\documentclass[aps,prl,twocolumn,notitlepage,superscriptaddress]{revtex4-1}

\usepackage{amsmath}
\usepackage{amssymb}
\usepackage{hyperref}
\usepackage{graphicx}
\usepackage[arrow, matrix, curve]{xy}
\usepackage{bbm}
\usepackage{bm}
\usepackage{yhmath}
\usepackage{color}
\usepackage{physics}
\usepackage{bbold}

\usepackage{color}

\renewcommand\vec[1]{\boldsymbol{\mathrm{#1}}}

\newcommand\diff{\mathrm{d}}

\usepackage[usenames,dvipsnames]{xcolor}
\hypersetup{colorlinks=true, linkcolor=BrickRed, urlcolor=blue!50!black, citecolor=blue!50!black}

\usepackage[normalem]{ulem}
\makeatletter
\newcommand\hide@visible[1]{%
  \bgroup\fboxsep=.3ex\colorbox{Gray}{begin hide}%
  #1\colorbox{Gray}{end hide}\egroup%
}
\newcommand\hide@hidden[1]{%
  \bgroup\fboxsep=.3ex\colorbox{Gray}{hidden text}%
}
\newcommand\hide@invisible[1]{}
\newcommand\makevisible{\let\hide\hide@visible}
\newcommand\makehidden{\let\hide\hide@hidden}
\newcommand\makeinvisible{\let\hide\hide@invisible}
\makeatother
\makehidden

\begin{document}

\title{Analytic Solution of an Active Brownian Particle in a Harmonic Well}

\author{Michele Caraglio}
\affiliation{Institut f\"ur Theoretische Physik, Universit\"at Innsbruck, Technikerstra{\ss}e~25/2,  A-6020 Innsbruck, Austria}

\author{Thomas Franosch}
\affiliation{Institut f\"ur Theoretische Physik, Universit\"at Innsbruck, Technikerstra{\ss}e~25/2,  A-6020 Innsbruck, Austria}
\email[]{thomas.franosch@uibk.ac.at}
%\thanks{}
 
\date{\today}

%\pacs{}
%02.50.Ey 	Stochastic processes
%02.60.Nm 	Integral and integrodifferential equations

%\keywords{Fokker-Planck equation}


\begin{abstract}
We provide an analytical solution for the time-dependent Fokker-Planck equation for a two-dimensional active Brownian particle trapped in an isotropic harmonic potential. Using the passive Brownian particle as basis states we show that the Fokker-Planck operator becomes lower diagonal, implying that the eigenvalues are unaffected by the activity. The propagator is then expressed as a combination of the equilibrium eigenstates with weights obeying exact iterative relations.
We show that for the low-order correlation functions, such as the positional autocorrelation function, the recursion terminates at finite order in the P{\'e}clet number allowing us to generate exact compact expressions and derive the velocity autocorrelation function and the time-dependent diffusion coefficient.
The nonmonotonic behavior of latter quantities serves as a fingerprint of the non-equilibrium dynamics.
\end{abstract}

\maketitle

It is hard to overstate the role of the harmonic oscillator is physics.
Being a paradigmatic model for waves and vibrational phenomena, it serves as a workhorse in both classical and quantum physics describing diverse phenomena such as springs, pendulums, molecular vibrations, acoustic oscillations, laser traps, electromagnetic fields in a cavity, and resonant electrical circuits, just to name a few~\cite{FeynmanLectures,SakuraiQM}.
Any  smooth potential can be approximated by a harmonic potential in the vicinity of a stable equilibrium point~\cite{FeynmanLectures} and even advanced tools like second quantization in quantum field theory have their roots in the mathematics of harmonic oscillations~\cite{Peskin1995}.%

Active matter and directed motion have come under the spotlight of several research communities, including biology~\cite{Berg2004,Lipowsky2005,Lauga2009,Julicher1997}, biomedicine~\cite{Naahidi2013,Liu2016,Henkes2020}, robotics~\cite{Cheang2014,Erkoc2019}, and statistical physics~\cite{Cates2012,Romanczuk2012,Marchetti2013,Chaudhuri2014,Pohl2014, Elgeti2015,Bechinger2016,Fodor2016,Falasco2016,Speck2016,Fodor2018,Vicsek1995,Toner1998}.
However, notwithstanding more than two decades of scientific efforts on self-propelled particles, some basic theoretical aspects have remained elusive since exactly solvable models of even single active particles are rare.
Generally, in external confining potentials, the steady-state probability distribution is not known analytically, with  the notable exceptions of active Brownian particles in channels~\cite{Wagner2017} or sedimenting in a gravitational field~\cite{Hermann2018}, and run-and-tumble particles in one dimension~\cite{Schnitzer1993,Tailleur2008,Tailleur2009,Malakar2018}. 
The complete characterization of the time-dependent probability distribution for a particle starting with certain initial conditions is even more challenging.
In this case, no analytical expressions are known for confining potentials, and in free space only solutions in the  Fourier domain have been provided for single active Brownian particle~\cite{Sevilla2015,Kurzthaler2016,Kurzthaler2017,Kurzthaler2018} and for run-and-tumble dynamics~\cite{Martens2012}.

The active Brownian particle (ABP) has become the minimal paradigm for self-propelled particles and it is already able to describe with a certain accuracy the properties of motion of a large fraction of existing microswimmers~\cite{Bechinger2016,Romanczuk2012}.
Such active particle can be trapped and monitored by optical~\cite{Ashkin1970,*Ashkin1980,*Ashkin1997} or acoustic~\cite{Takatori2016} tweezers which are well represented by harmonic potentials.
While simulations of ABPs in a harmonic trap can easily be performed by integrating the Langevin equations of motion, analytical progress is hindered because, despite the linearity of the restoring force, the problem remains nonlinear due to the constraint that the orientation can merely rotate. 
Recent significant advance has been achieved by Malakar \textit{et al.}~\cite{Malakar2020} for the stationary solution of the associated Fokker-Planck equation. 
They express the steady-state probability in the form of a power-series expansion in the P{\'e}clet number, a parameter indicating the relative importance of active motion compared to diffusion. However, the full time-dependent probability distribution of an ABP in a harmonic trap still remains elusive.

Here we show that, taking the eigenstates of the passive Brownian particle as an orthonormal basis and upon proper ordering of these states, the entire Fokker-Planck operator becomes lower diagonal.
This implies that not only the ground state but the entire eigenvalue spectrum of the Fokker-Planck operator remains unaltered when introducing the activity.
These surprising findings allows us to provide an exact expression for the probability propagator of an ABP in a two-dimensional harmonic well, thus going beyond existing theoretical approximations~\cite{Basu2018,Basu2019,Pototsky2012} and complementing numerical simulations, and experiments~\cite{Takatori2016,Dauchot2019}.
We also show that exact expressions of any moment or correlation function can be readily derived from our solution.
\medskip

%%%%%%%%%%%%%%%%%%%%%%%%%%%%%%%%%%%%%%%%%%%%%%%%%%%%%%%%%%%%%%%%

\textit{Model.}
We characterize the overdamped motion of a two-dimensional ABP in terms of the propagator $\mathbb{P}(\vec{r}, \vartheta, t | \vec{r}_0 , \vartheta_0)$ which is the probability to find the particle at position $\vec{r}$ and orientation $\vartheta$ at lag time $t$ given the initial position $\vec{r}_0$ and orientation $\vartheta_0$ at time $t=0$.
Its time evolution is provided by the Fokker-Planck equation~\cite{Risken1989}  
\begin{align} \label{eq:eom_propagator}
\partial_t\mathbb{P}  = 
 & \vec{\nabla} \cdot ( \mu k \vec{r} \mathbb{P}  ) + D \nabla^2 \mathbb{P} + D_{\text{rot}} \partial_\vartheta^2 \mathbb{P} -  v \vec{u} \cdot \vec{\nabla} \mathbb{P} \; ,
\end{align}
in short $\partial_t \mathbb{P} = \Omega\, \mathbb{P}$ with  $\Omega$  the Fokker-Planck operator and the formal solution of the propagator is thus 
$\mathbb{P}(\vec{r}, \vartheta, t | \vec{r}_0 , \vartheta_0) = e^{\Omega t} \delta(\vec{r}-\vec{r}_0) \delta(\vartheta-\vartheta_0) $.
The first term on the r.h.s. of Eq.~\eqref{eq:eom_propagator} describes the drift motion due to the harmonic potential $U(r) = k \vec{r}^2/2 $ with spring constant $k>0$, whereas $\mu$ is the mobility of the particle. 
The second term encodes the translational diffusion with diffusion coefficient $D$. 
The ratio $D/\mu = k_B T$ introduces an effective temperature which for a passive particle corresponds to the temperature of the solvent.
The rotational diffusion of the ABP is described by the third term with rotational diffusion coefficient $D_{\text{rot}}$, while the last term corresponds to the self-propulsion of the particle with fixed velocity $v$ along the orientation of the particle, $\vec{u} = (\cos\vartheta, \sin \vartheta)$.
In the case of a passive particle, $v=0$, 
the equilibrium distribution corresponds to the Boltzmann distribution $p^{\text{eq}}(\vec{r},\vartheta) \propto e^{-U(r)/k_B T}$, or, with proper normalization $\int \diff \vec{r} \diff \vartheta \, p^{\text{eq}}(\vec{r},\vartheta) = 1$,
\begin{align} \label{eq:equilibrium_distribution_passive_particle}
p^{\text{eq}}(\vec{r},\vartheta) %= \frac{1}{(2\pi)^2} \frac{\mu k}{ D} e^{ -\mu k r^2/2D } 
= \dfrac{\exp(-r^2/2 d^2)}{4 \pi^2 d^2} \; ,
\end{align}
where $d := \sqrt{k_B T/k}$ is the thermal oscillator length.
In particular, the translational and orientational degrees of freedom are decoupled.
\medskip

%%%%%%%%%%%%%%%%%%%%%%%%%%%%%%%%%%%%%%%%%%%%%%%%

\textit{Theory.}
The Fokker-Planck operator $\Omega$ in Eq.~\eqref{eq:eom_propagator} appears to be non-Hermitian already in equilibrium, $v=0$.
However, in this case it can be made manifestly Hermitian by a gauge transformation~\cite{Risken1989}.
Here we circumvent this detour and define a new operator $\mathcal{L}$ by splitting off the equilibrium density 
\begin{align} \label{eq:L_definition}
\Omega \left[  p^{\text{eq}}(\vec{r},\vartheta) \psi(\vec{r},\vartheta) \right]  =: p^{\text{eq}}(\vec{r},\vartheta) \mathcal{L} \psi(\vec{r},\vartheta) , 
\end{align}
where $\psi(\vec{r},\vartheta)$ is an arbitrary function depending on the coordinates $\vec{r}$ and $\vartheta$ only.
Then $\mathcal{L}$ can be naturally  decomposed 
\begin{align} \label{eq:decomposition_of_L}
 \mathcal{L} = \mathcal{L}_0 + \text{Pe} \, \mathcal{L}_1  ,
\end{align}
into an equilibrium contribution $\mathcal{L}_0$ and the non-equilibrium driving $\mathcal{L}_1$,  where $\text{Pe} := v d /D$ denotes the P{\'e}clet number and in the following will act as an expansion parameter.
In polar coordinates $\vec{r} = r (\cos\varphi, \sin\varphi)$  the equilibrium operator is expressed as
\begin{align} \label{eq:operator_L0}
\mathcal{L}_0\psi = \frac{D}{r} \partial_r \left( r\partial_r\psi\right) + \frac{D}{r^2} \partial_\varphi^2 \psi +
 D_{\text{rot}} \partial_\vartheta^2 \psi - \frac{D r}{d^2}  \partial_r  \psi  \; ,
\end{align}
%with the trap relaxation time $\tau := d^2/D $,
while the active part reads
\begin{align}\label{eq:operator_L1}
\mathcal{L}_1\psi 
% :=  - \frac{d}{\tau} \cos(\chi) \partial_r \psi - \frac{d}{\tau r}\sin(\chi) \partial_\varphi\psi + \frac{r d}{\tau} \cos (\chi) \psi \; . \\
 =  \frac{D}{d} \left[ -  \cos(\chi) \partial_r \psi - \frac{1}{ r}\sin(\chi) \partial_\varphi\psi + \frac{r}{d^2} \cos (\chi) \psi \right]  \; ,
\end{align}
where $\chi := \angle (\vec{u}, \vec{r}) = \vartheta-\varphi$ abbreviates the relative angle between orientation and position.

Then one readily shows that the equilibrium operator $\mathcal{L}_0$ is Hermitian, $\langle \phi | \mathcal{L}_0 \psi\rangle = \langle \mathcal{L}_0 \phi| \psi \rangle$, with respect to the 
Kubo scalar product
\begin{align}\label{eq:Kubo_scalar_product}
\langle \phi | \psi \rangle := \int \diff \vec{r}\! \int_0^{2\pi}\!\!\diff\vartheta \, p^{\text{eq}}(\vec{r},\vartheta)  \phi(\vec{r},\vartheta)^* \psi(\vec{r},\vartheta) \; , 
\end{align}
and correspondingly its eigenvalues are real and left and right eigenfunctions coincide. 
The solution the Hermitian eigenvalue problem of the equilibrium reference system,
\begin{align} \label{eq:eigenvalue_problem_LO}
\mathcal{L}_0 \psi =- \lambda \psi   \; ,
\end{align} 
is obtained by a separation ansatz following precisely the steps of the 2D isotropic harmonic oscillator in the quantum case~\cite{Pauli1973,Fluegge1999} augmented by the uncoupled orientational diffusion. 
Explicitly, the eigenfunctions read
\begin{align} \label{eq:eigenfunctions}
\psi_{n,\ell,j}(\vec{r},\vartheta) \!=\! \sqrt{\! \frac{n!}{(n \!+\! |\ell|)!}} 
  \left(\frac{r}{d \sqrt{2}} \right)^{\! |\ell|} \! \text{L}_n^{|\ell|} \! \left(\frac{r^2}{2 d^2} \! \right)
e^{i \ell \varphi } e^{i (j-\ell) \vartheta},
\end{align}
where $\text{L}_n^{|\ell|}(x)$ are the generalized Laguerre polynomials~\cite{NIST}. 
Here $n\in \mathbb{N}_0$ and $\ell, j \in \mathbb{Z}$.
The quantum numbers $(n,\ell, j)$ correspond to the 3 degrees of freedom $(r,\varphi,\vartheta)$ in polar coordinates. 
The associated eigenvalue is
\begin{align} \label{eq:eigenvalues}
\lambda_{n,\ell, j}  = \frac{1}{\tau} (2n + |\ell| ) + D_{\text{rot}} (j-\ell)^2  \; ,
\end{align}
with the trap relaxation time $\tau = d^2 /D =1 /\mu k$.

Since $\mathcal{L}_0$ is unchanged under rotations of  the position or the orientation of the particle it commutes with  the corresponding generators $L=-i \partial_\varphi$ and $S=-i \partial_\vartheta$ which, borrowing a quantum language, we refer to as `orbital momentum' and `spin'.
The eigenfunctions $\psi_{n,\ell, j}$ are simultaneous eigenfunctions to orbital momentum and spin with eigenvalues $\ell$ and $s:= j-\ell$.
For the active particle $\mathcal{L} = \mathcal{L}_0 + \text{Pe} \, \mathcal{L}_1$ remains invariant only under a simultaneous rotation of position and orientation, such that the total `angular momentum' $J = L + S$ is conserved. 
Hence, in the full problem $j$ will be still a good quantum number. 

Note that the eigenfunctions of the equilibrium reference system are orthonormalized with respect to the Kubo scalar product~\eqref{eq:Kubo_scalar_product}
\begin{align} \label{eq:orthogonality_Hilbert_space}
\braket{\psi_{n',\ell',j'} }{\psi_{n,\ell,j}}
= \delta_{j,j'} \delta_{\ell,\ell'} \delta_{n,n'}  \; ,
\end{align} 
and fulfill the completeness relation, 
\begin{align}\label{eq:completeness} 
p^{\text{eq}}(\vec{r},\vartheta) \sum_{n=0}^{\infty} \sum_{\ell=-\infty}^{\infty}
\sum_{j=-\infty}^{\infty}
& \psi_{n,\ell,j}(\vec{r},\vartheta)   \psi_{n,\ell, j}(\vec{r}_0, \vartheta_0)^* \nonumber \\
  & = \delta(\vec{r}-\vec{r}_0) \delta(\vartheta-\vartheta_0) \; . 
\end{align}

\begin{figure*}[t!]
\centering
\includegraphics[scale=1]{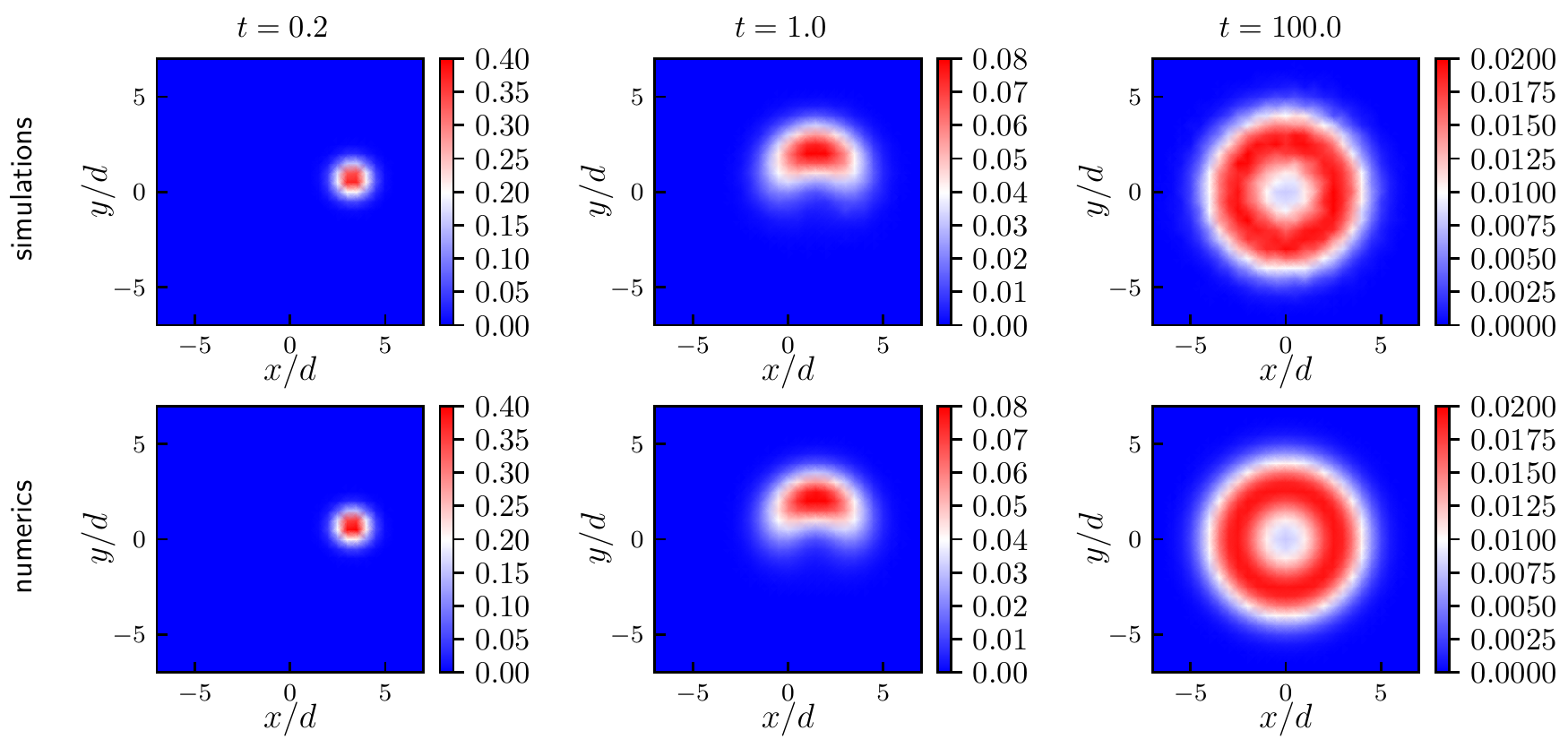}
\caption{Spatial probability distribution at different times $t$ starting with initial condition $r_0 = 4d$, $\varphi_0 = 0$, and $\vartheta_0 = \pi/2$. Comparison between simulations and numerics for $\text{Pe} = 4$ and $D_{\text{rot}} \tau=0.8$.
For the simulations, statistics has been collected from $2\cdot 10^5$ independent realizations of the process.\label{fig:spatial_probability}}
\end{figure*}

Moving our attention back to the full problem for an active particle, the formal  expression of the propagator allows us to write 
\begin{align}\label{eq:solution_propagator}
\mathbb{P} &(\vec{r}, \vartheta, t | \vec{r}_0 , \vartheta_0) = e^{\Omega t} \delta(\vec{r}-\vec{r}_0) \delta(\vartheta - \vartheta_0) \nonumber \\
& = p^{\text{eq}}(\vec{r},\vartheta) \sum_{n,\ell,j} \left\lbrace e^{\mathcal{L} t} \psi_{n,\ell,j}(\vec{r},\vartheta)  \right\rbrace \psi_{n,\ell,j}(\vec{r}_0,\vartheta_0)^* \nonumber \\
& = p^{\text{eq}}(\vec{r},\vartheta) \sum_{n,\ell,j} \bra{\vec{r}\vartheta} e^{\mathcal{L} t} \ket{ \psi_{n,\ell,j}} \braket{\psi_{n,\ell,j}}{\vec{r}_0 \vartheta_0}
\; ,
\end{align}
where, going from the first to the second line, we used Eqs.~\eqref{eq:completeness} and~\eqref{eq:L_definition} and, in the third line, we rely on Dirac's bra-ket notation where the isomorphism between $\ket{\psi}$ and $\psi(\vec{r},\vartheta)$ is made explicit by introducing generalized position/orientation states $|\vec{r} \vartheta\rangle$ such that $\psi(\vec{r},\vartheta) = \langle \vec{r}\vartheta |\psi \rangle$~\cite{note1}.
Then, exploiting twice the identity relation
\begin{align}\label{eq:identity}
\sum_{n,\ell,j} \ket{\psi_{n,\ell,j}}\bra{\psi_{n,\ell,j}} = \mathbb{1} \; ,
\end{align}
Eq.~\eqref{eq:solution_propagator} can be finally recast in
\begin{align}\label{eq:solution_propagator_withMs}
\mathbb{P} &(\vec{r}, \vartheta, t | \vec{r}_0 , \vartheta_0)  \nonumber \\
& =  p^{\text{eq}}(\vec{r},\vartheta)  \sum_{n,\ell,j}  M_{n,\ell,j} (\vec{r}_0,\vartheta_0,t) \, \psi_{n,\ell,j}(\vec{r},\vartheta)
\; ,
\end{align}
with
\begin{align} \label{eq:Ms}
M_{n,\ell,j} (\vec{r}_0,\vartheta_0,t) := \bra{\psi_{n,\ell,j}} e^{\mathcal{L} t} \ket{\vec{r}_0\vartheta_0} \; .
\end{align}
Note that the functions $M_{n,\ell,j} (\vec{r}_0,\vartheta_0,t)$ depend only on time $t$ and on the initial conditions $(\vec{r}_0,\vartheta_0)$, which greatly simplifies the numerical implementation.

To make further progress we rely on the renowned Dyson equation, familiar from quantum theory~\cite{SakuraiQM}, for the time evolution operator
\begin{align} \label{eq:Dyson_series}
e^{\mathcal{L} t} = e^{\mathcal{L}_0 t} + \text{Pe} \int_0^t \diff s \, e^{\mathcal{L}_0 (t-s)} \, \mathcal{L}_1 \, e^{\mathcal{L} s} \; ,
\end{align}
which can be inserted in Eq.~\eqref{eq:Ms}, together with the identity~\eqref{eq:identity}, to obtain a useful integral relation for the functions $M$ appearing in the propagator
\begin{align} \label{eq:recursion_Ms}
& M_{n,\ell,j} (\vec{r}_0,\vartheta_0,t) =  \; e^{-\lambda_{n,\ell,j} t} \braket{\psi_{n,\ell,j}}{\vec{r}_0\vartheta_0} + \nonumber \\
& \qquad + \! \text{Pe} \! \int_0^t \!\! \diff s  \, \bigg[ e^{- \! \lambda_{n,\ell,j} (t-s)} \!\! \nonumber \\
& \times \sum_{n',\ell',j'} \! \bra{\psi_{n,\ell,j}} \! \mathcal{L}_1 \! \ket{\psi_{n',\ell',j'}} 
M_{n',\ell',j'}(\vec{r}_0,\! \vartheta_0,\!s) \bigg] \, .
\end{align}

For active particles, $\text{Pe}>0$, the operator $\mathcal{L}_1$ introduces couplings between the eigenstates $\ket{\psi_{n, \ell, j}}$. 
Starting from Eqs.~\eqref{eq:operator_L1} and~\eqref{eq:eigenfunctions}, one readily obtains (see also Ref.~\cite{Malakar2020} for a comparison to the steady-state solution)
\begin{align} \label{eq:L1_action}
 & \mathcal{L}_1 \ket{\psi_{n,\ell,j}}  =  \nonumber \\
 &  \dfrac{1}{\sqrt{2} \tau} \! \left\lbrace \!\!\!
\begin{array}{lr}
\sqrt{n \!+\! \ell \!+\! 1} \, \ket{\psi_{n,\ell+1,j}} \!-\! \sqrt{n \!+\! 1} \, \ket{\psi_{n+1,\ell-1,j}} & \mbox{ if } \ell \!>\! 0 ,\\ \\
\sqrt{n \!+\! 1} \, \ket{\psi_{n,\ell+1,j}} + \sqrt{n \!+\! 1} \, \ket{\psi_{n,\ell-1,j}} & \mbox{ if } \ell \!=\! 0 ,\\ \\
\sqrt{n \!-\! \ell \!+\! 1} \, \ket{\psi_{n,\ell-1,j}} \!-\! \sqrt{n \!+\! 1} \, \ket{\psi_{n+1,\ell+1,j}} & \mbox{ if } \ell \!<\! 0 .
\end{array}
\right.
\end{align}
As anticipated, the action of the operator $\mathcal{L}_1$ does not modify the quantum number $j$.
Furthermore, its nature is such that, when applied to $\ket{\psi_{n,\ell,j}}$, $n$ never decreases and either $|\ell|$ or $n$ increases by $1$.
Thus, if the eigenstates are ordered according to the value of $2n+|\ell|$, $\mathcal{L}_1$ and its powers $(\mathcal{L}_1)^q$ with $q>1$, are strictly lower diagonal matrices in the eigenbasis of $\mathcal{L}_0$.

A first consequence is that, surprisingly, the entire spectrum of the full problem remains unaltered with respect to the reference passive system.
Furthermore, these two properties allow calculating the $M$'s exactly in a iterative scheme which starts from $M_{0,0,j}(\vec{r}_0,\vartheta_0,t) = e^{-\lambda_{0,0,j} t}\psi_{0,0,j}(\vec{r}_0,\vartheta_0)^*$ and progressively builds the $M_{n,\ell,j}$'s such that $2n+|\ell|=q$ once the $M_{n,\ell,j}$'s such that $2n+|\ell|=q-1$ are known.
Thus, for a given $n$ and $\ell$, the corresponding $M_{n,\ell,j}$'s result in a linear combination of a finite number of eigenfunctions. 
For example
\begin{align}
M_{0,1,j} & (\vec{r}_0,\vartheta_0,t) = e^{-\lambda_{0,1,j} t} \psi_{0,1,j}(\vec{r}_0,\vartheta_0)^* \nonumber \\
& + \dfrac{\text{Pe}}{\sqrt{2} \tau} \, \dfrac{e^{-\lambda_{0,0,j} t}-e^{-\lambda_{0,1,j} t}}{\lambda_{0,1,j} - \lambda_{0,0,j}} \psi_{0,0,j}(\vec{r}_0,\vartheta_0)^* \; .
\end{align}
Few more explicit expression for low values of $2n+|\ell|$ are reported in the Supplemental Material (SM)~\cite{supplement_material}.
Intuitively, the analytical evaluation of these functions becomes quickly tedious with increasing $2n+|\ell|$.
However, they are efficiently computed numerically exploiting an integrated version of Eq.~\eqref{eq:recursion_Ms} which is also reported in the SM~\cite{supplement_material}.
\medskip

%%%%%%%%%%%%%%%%%%%%%%%%%%%%%%%%%%%%%%%%%%%%%%%%

\textit{Results.}
To corroborate our findings, we benchmark several observables that can be evaluated by exploiting Eq.~\eqref{eq:solution_propagator_withMs} against their analog obtained by directly solving the Langevin equation of motion.
As a first example, we report in Fig.~\ref{fig:spatial_probability} the time evolution of the spatial probability distribution starting from some given initial condition.
As a side note, we stress here that, in the case reported in Fig.~\ref{fig:spatial_probability}, collecting enough statistics for this observable from Langevin simulations is about 20 times slower than obtaining the result from the numerics.

\begin{figure}
\centering
\includegraphics[scale=1]{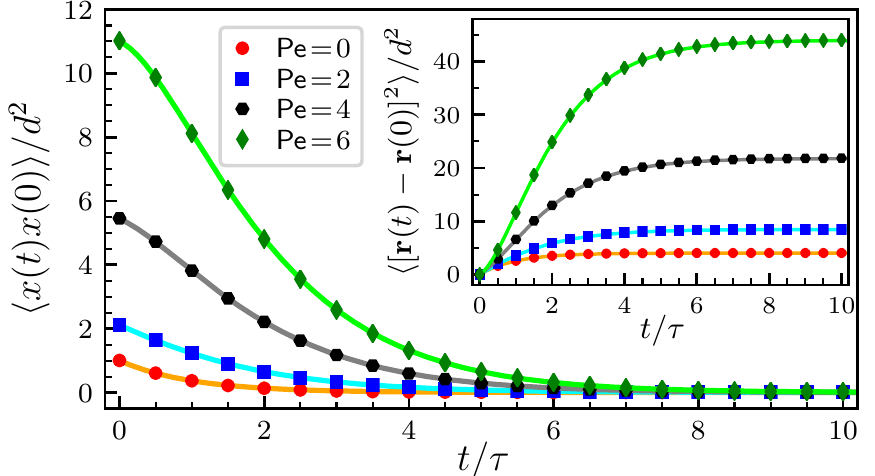} 
\caption{Positional autocorrelation function (main panel) and mean square displacement (inset) vs. lag time $t$. Comparison between simulations (symbols) and analytical results (lines) for $D_{\text{rot}} \tau=0.8$ at various values of P{\'e}clet number $\text{Pe}$.\label{fig:PAF_MSD}}
\end{figure}

Equation~\eqref{eq:solution_propagator_withMs} also easily allows calculating some paradigmatic moments and correlation functions.
While exact computation of moments starting from a given initial condition has recently been discussed~\cite{Chaudhuri2021}, our approach provides an alternative and simple way for obtaining them in terms of the functions $M$'s.
In fact, given the orthogonality relation~\eqref{eq:orthogonality_Hilbert_space}, integration over positional and directional degrees of freedom truncates the infinite series appearing in the propagator such that moments result in a combination of a finite number of $M$ functions.
For instance
\begin{align}
\langle [\vec{r}&(t)]^2  \rangle_{\vec{r}_0,\vartheta_0} = \int_0^{\infty} \! \diff r \int_0^{2\pi} \! \diff \varphi \int_0^{2\pi} \! \diff \vartheta \; r^3 \, \mathbb{P} (\vec{r}, \vartheta, t | \vec{r}_0 , \vartheta_0) \nonumber \\
& = 2 d^2 \left[  M_{0,0,0}(\vec{r}_0,\vartheta_0,t) - M_{1,0,0}(\vec{r}_0,\vartheta_0,t) \right] \; .
\end{align}

Furthermore, we present here, for the first time, exact analytical expressions also  for moments and correlation functions averaged over the initial conditions, which are the genuine quantities directly accessible in experiments.
In particular, the positional autocorrelation function (PAF)
%\cite{Franosch2011} 
reads
\begin{align} \label{eq:PAF}
\langle x(t) x(0) \rangle = d^2 \! \left[ e^{-t/\tau} \!-\! \dfrac{\text{Pe}^2}{2} \dfrac{  D_{\text{rot}} \tau  e^{-t/\tau} \!-\! e^{-D_{\text{rot}}t}}{1-( D_{\text{rot}} \tau)^2} \right] \; ,
\end{align}
while the mean square displacement becomes
\begin{align} \label{eq:MSD}
\langle [\vec{r}(t)-\vec{r}(0)]^2 \rangle =  4d^2 \big( 1 - &  e^{-t/\tau} \big) 
\nonumber \\
+ 2 \, \text{Pe}^2 d^2 \left[  \dfrac{1-e^{- t/\tau}}{1 +  D_{\text{rot}}\tau } \right. + & \left. \dfrac{e^{-t/\tau}-e^{- D_{\text{rot}} t}}{1 - ( D_{\text{rot}}\tau)^2 } \right] \; .
\end{align}
See SM~\cite{supplement_material} for details and Fig.~\ref{fig:PAF_MSD} for a comparison with Langevin simulations.
Interestingly, the effect of the activity on the previous quantities is characterized only by terms of second order in the P{\'e}clet number.

\begin{figure}
\centering
\includegraphics[scale=1]{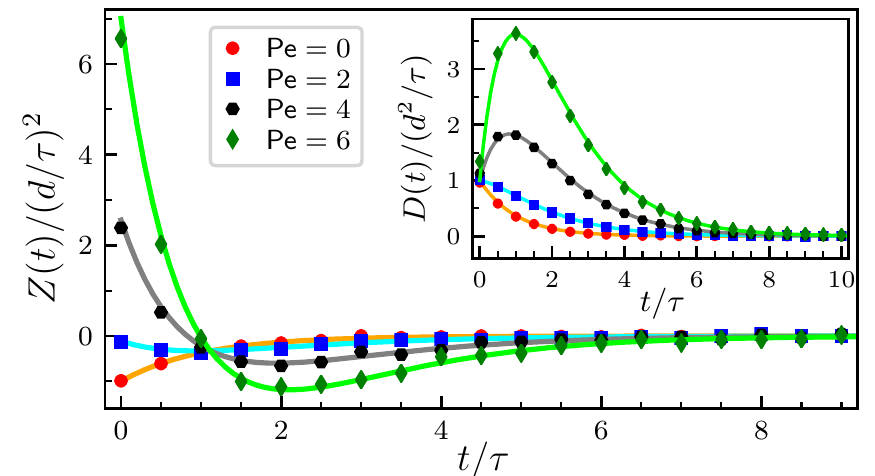} 
\caption{Velocity autocorrelation function (main panel) and time-dependent diffusion coefficient (inset) vs. lag time $t$. Comparison between simulations (symbols) and analytical results (lines) for $D_{\text{rot}} \tau=0.8$ at various P{\'e}clet numbers $\text{Pe}$. \label{fig:Z_D}}
\end{figure}

The non-trivial contribution of the activity to the dynamics becomes even more evident when considering the velocity autocorrelation function (VACF)
%~\cite{Franosch2011} 
which is defined for $t>0$ as
\begin{align}
Z (t) := -\dfrac{\diff^2}{\diff t^2} \langle x(t) x(0) \rangle \; .
\end{align}
In equilibrium, any correlation function is a \textit{completely monotone function}, i.e. the correlation function and all of its derivatives decay monotonically~\cite{Feller1970,Leitmann2017}.
We thus expect a negative and strictly increasing behavior of the passive VACF.
In contrast, the VACF of ABPs displays a nonmonotonic behavior which becomes more pronounced with the activity and with a minimal value whose position increases with the P{\'e}clet number, see Fig.~\ref{fig:Z_D}.
Furthermore, if the activity contribution is strong enough, the VACF becomes positive for small times.
Similar observations hold also for the time-dependent diffusion coefficient (inset of Fig.~\ref{fig:Z_D}) 
\begin{align}
D (t) := \dfrac{1}{4} \dfrac{\diff}{\diff t} \langle [\vec{r}(t)-\vec{r}(0)]^2 \rangle \; .
\end{align}
%which is reported in the inset of Fig.~\ref{fig:Z_D}.
%\medskip

%%%%%%%%%%%%%%%%%%%%%%%%%%%%%%%%%%%%%%%%%%%%%%%%

\textit{Conclusions.} 
We have derived and illustrated an exact series solution for the probability propagator of an ABP confined to a two-dimensional harmonic trap.
Such a solution is obtained by dealing with the activity of the particle in a perturbative approach, which is feasible because the Fokker-Planck operator becomes lower diagonal when the eigenstates of the passive reference system are taken as a basis and properly sorted.
This surprising property allows us to express the propagator as a combination of the unperturbed eigenfunctions weighted by factors that depend only on time and the initial conditions and that can efficiently be computed in an exact iterative scheme.
This property also implies that not only the ground state but the entire eigenvalue spectrum remains unaltered when introducing the activity.
Consequently, the propagator can also be expressed in terms of the perturbed left and right eigenfunctions~\cite{note2} multiplied by an exponentially decaying factor with a rate given by the corresponding unperturbed eigenvalue.
The propagator also provides the steady-state distribution in the long-time limit and, in this regime, our expression, becomes equivalent to that given by Malakar \textit{et al.}~\cite{Malakar2020}.
From our solution, paradigmatic moments and correlation functions are then readily obtained and show expressions terminating at finite order in the P{\'e}clet number.
The VACF and the time-dependent diffusion coefficient of ABPs derived from the PAF and the mean square displacement, display a nonmonotonic behavior which truly reveal their non-equilibrium character.

Our work provides a definitive and unifying framework encompassing previous theoretical results~\cite{Malakar2020,Basu2018,Basu2019,Pototsky2012,Chaudhuri2021,Dauchot2019} on the behavior of a single ABP in a harmonic trap and sheds light on the relationships among them.
Beyond its fundamental relevance, this is particularly important in view of the fact that the harmonic potential is an approximation of any potential in the vicinity of a stable point.
Thus, being able to exactly describe the dynamics of an ABP in a harmonic well is a first step towards a deeper understanding of their behavior in more complicated potentials and, as such, may have an impact on several applications as first-passage time~\cite{Geiseler2016,Woillez2019,Caprini2019,Tejedor2012} and target-search~\cite{Tejedor2012,Volpe2017,Zanovello2021,Zanovello2021b} problems, just to mention a few.
Not only our findings can be generalized to chiral ABP~\cite{vanTeeffelen2008} by adding a drift term to the dynamics of the orientation of the particle,
but they may also serve as a starting point to solve the dynamics of active molecules~\cite{Babel2016,Kuchler2016,Lowen2018} and active polymers~\cite{Kaiser2015,MartinGomez2019} in which the constitutive beads are bonded via spring-like potentials.
For instance, the case of an active dumbbell~\cite{Lowen2018,Winkler2016} composed of an ABP and a passive Brownian particle can be mapped to our model.
Furthermore, a careful investigation of the limit of vanishing potential could shed new light also on the behavior of ABPs in free space~\cite{Sevilla2015,Kurzthaler2016,Kurzthaler2017,Kurzthaler2018}.
Fitting to our analytical expressions moments and correlation functions measured in experiments of Janus particles~\cite{Howse2007,Jiang2010} trapped by optical~\cite{Ashkin1970,*Ashkin1980,*Ashkin1997} or acoustic~\cite{Takatori2016} tweezers may provide a robust method to determine their P{\'e}clet number and rotational diffusion coefficient.
Finally, the ABP in a harmonic well may be used as a toy model to illustrate generic results in non-equilibrium thermodynamics such as trade-off relations between between speed, uncertainty, and dissipation~\cite{Gingrich2016,Pietzonka2016,Neri2022,Shiraishi2018}.

A last remark is in order: 
The harmonic oscillator is special in many ways already at the level of a passive particle.
The degeneracy of the eigenvalue spectrum in the quantum case is connected to a higher symmetry $\text{SU}(2)$ beyond rotational symmetry $\text{SO}(2)$~\cite{Fradkin1965,*Fradkin1967,note3} which by Noether's theorem~\cite{Olver2000} implies the existence of a conserved quantity known as the Fradkin tensor (the analog of the Runge-Lenz vector in the Kepler problem).
These consideration transfer to the passive particle and one readily finds the corresponding Fradkin tensor.
Generally, any perturbation reduces the symmetry and the levels are anticipated to split, compare e.g.\@ to the Stark effect~\cite{SakuraiQM}. 
However, our analysis reveals that the spectrum is unaffected by the activity, in particular, the degeneracy of the spectrum is not lifted.
One is tempted to argue that also the activity reflects at least a variant of the $\text{SU}(2)$ symmetry. 
However, simple guesses to generalize the Fradkin tensor to the case of an active particle fail, and Noether's theorem does not directly apply since the Fokker-Planck equation, Eq.~\eqref{eq:eom_propagator}, does not derive from a variational principle in the non-equilibrium case. 
Therefore pinpointing down the origin of the degeneracy in the active case remains a challenge for the future.

\begin{acknowledgements}
\textit{Acknowledgments.}
We thank Christina Kurzthaler for constructive criticism on the manuscript. TF acknowledges funding by FWF: P 35580-N.
\end{acknowledgements}

%%%%%%%%%%%%%%%%%%%%%%%%%%%%%%%%%%%%%%%%%%%%%%%%%%%%%%%%%%%%%%%%
%\bibliographystyle{apsrev4-1-title}
%\bibliography{references}

\begin{thebibliography}{79}
\bibitem [{not({\natexlab{a}})}]{note1}%
  \BibitemOpen
  \href@noop {} {}\bibinfo {note} {Note that here and in the following, the
  sandwich notation is used with the specification that, being $\mathcal{L}$ a
  non-Hermitian operator, it only acts on the right.}\BibitemShut {Stop}%
\bibitem [{not({\natexlab{b}})}]{note2}%
  \BibitemOpen
  \href@noop {} {}\bibinfo {note} {The perturbed left and right eigenfunctions
  can be easily computed by direct diagonalization of the matrix form of the
  operator $\mathcal{L}$.}\BibitemShut {Stop}%
\bibitem [{not({\natexlab{c}})}]{note3}%
  \BibitemOpen
  \href@noop {} {}\bibinfo {note} {Note that we are in two dimensions and
  therefore the symmetries are lower with respect to the original work by
  Fradkin.}\BibitemShut {Stop}%
\end{thebibliography}

\newpage
\onecolumngrid

\begin{center}
\begin{LARGE}
\textbf{Supplemental Material for}
\end{LARGE}
\\
\begin{Large}
``Analytic Solution of an Active Brownian Particle in a Harmonic Well''
\end{Large}
\end{center}

\bigskip

%%%%%%%%%% Merge with supplemental materials %%%%%%%%%%
%%%%%%%%%% Prefix a "S" to all equations, figures, tables and reset the counter %%%%%%%%%%
\setcounter{equation}{0}
\setcounter{figure}{0}
\setcounter{table}{0}
\setcounter{page}{1}
\makeatletter
\renewcommand{\theequation}{S\arabic{equation}}
\renewcommand{\thefigure}{S\arabic{figure}}
\renewcommand{\bibnumfmt}[1]{[S#1]}
\renewcommand{\citenumfont}[1]{S#1}

\begin{center}
{\large \textbf{Efficient computation of the functions $M_{n,\ell,j} (\vec{r}_0,\vartheta_0,t)$}}
\end{center}

As reported in the main text, the functions $M_{n,\ell,j} (\vec{r}_0,\vartheta_0,t)$ appearing in the expression of the propagator obey the following useful recursive relation
\begin{align} \label{eq:recursion_Ms_sup}
M_{n,\ell,j} (\vec{r}_0,\vartheta_0,t) = & \; e^{-\lambda_{n,\ell,j} t} \braket{\psi_{n,\ell,j}}{\vec{r}_0\vartheta_0} + \text{Pe} \int_0^t \diff s  \, e^{-\lambda_{n,\ell,j} (t-s)} \bra{\psi_{n,\ell,j}}  \mathcal{L}_1 \ket{\psi_{n',\ell',j'}}  
M_{n',\ell',j'}(\vec{r}_0,\vartheta_0,s) \nonumber \\
= & \; e^{-\lambda_{n,\ell,j} t} \psi_{n,\ell,j}(\vec{r}_0,\vartheta_0)^* \nonumber \\
+
\dfrac{\text{Pe}}{\sqrt{2} \, \tau} \int_0^t \diff s  &\, e^{-\lambda_{n,\ell,j} (t-s)} \left\lbrace 
\begin{array}{lr}
\sqrt{n \!+\! \ell} \, M_{n,\ell-1,j} (\vec{r}_0,\vartheta_0,s)
-\sqrt{n} \, M_{n-1,\ell+1,j} (\vec{r}_0,\vartheta_0,s)
& \qquad  \mbox{for } \ell>0
\; , \\
\, \\
-\sqrt{n} \, M_{n-1,\ell-1,j} (\vec{r}_0,\vartheta_0,s)
-\sqrt{n} \, M_{n-1,\ell+1,j} (\vec{r}_0,\vartheta_0,s)  & \qquad \mbox{for } \ell=0
\; , \\
\, \\
-\sqrt{n} \, M_{n-1,\ell-1,j} (\vec{r}_0,\vartheta_0,s)
+\sqrt{n \!-\! \ell} \, M_{n,\ell+1,j} (\vec{r}_0,\vartheta_0,s) & \qquad  \mbox{for } \ell<0
\; ,
\end{array}
\right. 
\end{align}
where in the second line we made explicit the action of the operator $\mathcal{L}_1$.
Writing explicitly few of these functions
\begin{align}
M_{0,0,j}(\vec{r}_0,\vartheta_0,t) = e^{-\lambda_{0,0,j} t}\psi_{0,0,j}(\vec{r}_0,\vartheta_0)^* \; ,
\end{align}
\begin{align}
M_{0,1,j} (\vec{r}_0,\vartheta_0,t) = e^{-\lambda_{0,1,j} t} \psi_{0,1,j}(\vec{r}_0,\vartheta_0)^* + \dfrac{\text{Pe}}{\sqrt{2} \, \tau} \, \dfrac{e^{-\lambda_{0,0,j} t}-e^{-\lambda_{0,1,j} t}}{\lambda_{0,1,j} - \lambda_{0,0,j}} \psi_{0,0,j}(\vec{r}_0,\vartheta_0)^* \; ,
\end{align}
\begin{align}
M_{0,2,j} & (\vec{r}_0,\vartheta_0,t) = e^{-\lambda_{0,2,j} t} \psi_{0,2,j}(\vec{r}_0,\vartheta_0)^* + \dfrac{\text{Pe}}{\tau} \, \dfrac{e^{-\lambda_{0,1,j} t}-e^{-\lambda_{0,2,j} t}}{\lambda_{0,2,j} - \lambda_{0,1,j}} \psi_{0,1,j}(\vec{r}_0,\vartheta_0)^* \nonumber \\
& +  \dfrac{\text{Pe}^2}{\sqrt{2} \, \tau^2} \,  \dfrac{1}{\lambda_{0,1,j} - \lambda_{0,0,j}} \Big[ 
\dfrac{ e^{-\lambda_{0,0,j} t}-e^{-\lambda_{0,2,j} t}}{\lambda_{0,2,j} - \lambda_{0,0,j}} - \dfrac{ e^{-\lambda_{0,1,j} t}-e^{-\lambda_{0,2,j}t}  }{\lambda_{0,2,j} - \lambda_{0,1,j}}  
\Big] \psi_{0,0,j}(\vec{r}_0,\vartheta_0)^*
\; ,
\end{align}
\begin{align}
M_{1,0,j} & (\vec{r}_0,\vartheta_0,t)  =  e^{-\lambda_{1,0,j} t} \psi_{1,0,j}(\vec{r}_0,\vartheta_0)^*  \nonumber \\
& -  \dfrac{\text{Pe}}{\sqrt{2}\, \tau} \, \dfrac{e^{-\lambda_{0,1,j} t}-e^{-\lambda_{1,0,j} t}}{\lambda_{1,0,j} - \lambda_{0,1,j}}  \psi_{0,1,j}(\vec{r}_0,\vartheta_0)^* -  \dfrac{\text{Pe}}{\sqrt{2} \, \tau} \, \dfrac{e^{-\lambda_{0,-1,j} t}-e^{-\lambda_{1,0,j} t}}{\lambda_{1,0,j} - \lambda_{0,-1,j}}  \psi_{0,-1,j}(\vec{r}_0,\vartheta_0)^* \nonumber \\
&  -  \dfrac{\text{Pe}^2}{2 \tau^2} \,  \dfrac{1}{\lambda_{0,1,j} - \lambda_{0,0,j}} \Big[ 
\dfrac{ e^{-\lambda_{0,0,j} t}-e^{-\lambda_{1,0,j} t}}{\lambda_{1,0,j} - \lambda_{0,0,j}} - \dfrac{ e^{-\lambda_{0,1,j} t}-e^{-\lambda_{1,0,j}t}  }{\lambda_{1,0,j} - \lambda_{0,1,j}}  
\Big]  \psi_{0,0,j}(\vec{r}_0,\vartheta_0)^*  \nonumber \\
&  - \dfrac{\text{Pe}^2}{2 \tau^2} \,  \dfrac{1}{\lambda_{0,-1,j} - \lambda_{0,0,j}} \Big[ 
\dfrac{ e^{-\lambda_{0,0,j} t}-e^{-\lambda_{1,0,j} t}}{\lambda_{1,0,j} - \lambda_{0,0,j}} - \dfrac{ e^{-\lambda_{0,-1,j} t}-e^{-\lambda_{1,0,j}t}  }{\lambda_{1,0,j} - \lambda_{0,-1,j}}  
\Big] \psi_{0,0,j}(\vec{r}_0,\vartheta_0)^* \; ,
\end{align}

As one easily realizes from the previous expressions, the time dependence can be separated from the position dependence as
\begin{align}
M_{n,\ell,j} (\vec{r}_0,\vartheta_0,t) = \sum_{q=0}^{2n+|\ell|} \! \left( \dfrac{\text{Pe}}{\sqrt{2}\, \tau}\right)^{\!q} \, \sum_{p=0}^q C_{n,\ell,j}^{(q,p)}(t) \; \psi_{n^{\star},\ell^{\star},j}(\vec{r}_0,\vartheta_0)^*
\end{align}
with 
$\ell^{\star} = \ell+q-2p$,
$n^{\star} = n - (q-|\ell|+|\ell^{\star}|)/2$ and
\begin{align} \label{eq:recursion_Cs_sup}
C_{n,\ell,j} ^{(q,p)}(t) = 
\int_0^t \diff s  &\, e^{-\lambda_{n,\ell,j} (t-s)} \left\lbrace 
\begin{array}{lr}
\sqrt{n \!+\! \ell} \, C_{n,\ell-1,j}^{(q-1,p-1)} (s)
-\sqrt{n} \, C_{n-1,\ell+1,j}^{(q-1,p)} (s)
& \qquad  \mbox{for } \ell>0
\; , \\
\, \\
-\sqrt{n} \, C_{n-1,\ell-1,j}^{(q-1,p-1)} (s)
-\sqrt{n} \, C_{n-1,\ell+1,j}^{(q-1,p)} (s)  & \qquad \mbox{for } \ell=0
\; , \\
\, \\
-\sqrt{n} \, C_{n-1,\ell-1,j}^{(q-1,p-1)} (s)
+\sqrt{n \!-\! \ell} \, C_{n,\ell+1,j}^{(q-1,p)} (s) & \qquad  \mbox{for } \ell<0
\; ,
\end{array}
\right. 
\end{align}
The previous iterative scheme is complemented by the boundary conditions
$C_{n,\ell,j} ^{(0,0)}(t) = e^{-\lambda_{n,\ell,j} t}$ and $C_{n,\ell,j} ^{(q,p)}(t) = 0$ if $n<1$ or $q<0$ or $p<0$ or $p>q$, and can be rewritten in the more convenient form (from the numerical evaluation point of view) as
\begin{align}
C_{n,\ell,j} ^{(q,p)}(t) \!=\! 
\dfrac{1}{\lambda_{n,\ell,j} \!-\! \lambda_{n^{\star}\!,\ell^{\star}\!,j}}
\left\lbrace \!\!
\begin{array}{lr}
\sqrt{n \!+\! \ell} \, C_{n,\ell-1,j}^{(q\!-\!1,p\!-\!1)} \! (t)
\!-\! F^p_{n^{\star}\!,\ell^{\star}} C_{n,\ell,j}^{(q-1,p-1)} \!(t)
\!-\!\sqrt{n} \, C_{n-1,\ell+1,j}^{(q-1,p)} \!(t)
\!-\! F^0_{n^{\star}\!,\ell^{\star}} C_{n,\ell,j}^{(q\!-\!1,p)} \!(t)
& \;  \mbox{if } \ell\!>\!0
\\
\, \\
-\sqrt{n} \, C_{n-1,\ell-1,j}^{(q\!-\!1,p\!-\!1)} \!(t)
\!-\! F^p_{n^{\star}\!,\ell^{\star}} C_{n,\ell,j}^{(q-1,p-1)} \!(t)
\!-\!\sqrt{n} \, C_{n-1,\ell+1,j}^{(q-1,p)} \!(t)
\!-\! F^0_{n^{\star}\!,\ell^{\star}} C_{n,\ell,j}^{(q\!-\!1,p)} \!(t)
& \; \mbox{if } \ell\!=\!0 \\
\, \\
-\sqrt{n} \, C_{n-1,\ell-1,j}^{(q\!-\!1,p\!-\!1)} \!(t)
\!-\! F^p_{n^{\star}\!,\ell^{\star}} C_{n,\ell,j}^{(q-1,p-1)} \!(t)
\!+\!\sqrt{n\!-\!\ell} \, C_{n,\ell+1,j}^{(q-1,p)} \!(t)
\!-\! F^0_{n^{\star}\!,\ell^{\star}} C_{n,\ell,j}^{(q\!-\!1,p)} \!(t) & \;  \mbox{if } \ell\!<\!0
\end{array}
\right. 
\end{align}
with
\begin{align}
F^p_{n^{\star}\!,\ell^{\star}} = \left\lbrace \!\!
\begin{array}{lr}
\sqrt{n^{\star} \!+\! \ell^{\star} \!+\! 1}
& \;  \mbox{if } \ell^{\star} \!\geq\!0
\\
\, \\
-\sqrt{n^{\star} \!+\! 1}
& \; \mbox{if } \ell^{\star}  \!<\!0 \\
\end{array}
\right. 
\qquad
\qquad
F^0_{n^{\star}\!,\ell^{\star}} = \left\lbrace \!\!
\begin{array}{lr}
-\sqrt{n^{\star} \!+\! 1}
& \;  \mbox{if } \ell^{\star} \!>\!0
\\
\, \\
\sqrt{n^{\star}  \!-\! \ell^{\star} \!+\! 1}
& \; \mbox{if } \ell^{\star}  \!\leq\!0 \\
\end{array}
\right. 
\end{align}
The previous scheme is particularly efficient to implement from the numerical point of view and allows for an evaluation of the propagator which is computationally faster than exploiting direct diagonalization of the matrix form of the operator $\mathcal{L}$ to obtain the left and right perturbed eigenfunctions (see main text).
The apparent singular behavior for $\lambda_{n,\ell,j} \rightarrow \lambda_{n^{\star}\!,\ell^{\star}\!,j}$ can also be solved through an iterative scheme:
\begin{align}
\lim_{\lambda_{n,\ell,j} \rightarrow \lambda_{n^{\star}\!,\ell^{\star}\!,j}}   C_{n,\ell,j} ^{(q,p)}(t) = D_{n^{\star},\ell^{\star},j}^{(q,p)}(t)
\end{align}
with
\begin{align} \label{eq_iterations_of_D_sup}
D_{n^{\star},\ell^{\star},j}^{(q,p)}(t) = 
\left\lbrace \!\!
\begin{array}{lr}
\sqrt{\tilde{n} \!+\! \tilde{\ell}} \dfrac{
C_{\tilde{n},\tilde{\ell}-1,j} ^{(q-1,p-1)}(t) - D_{n^{\star},\ell^{\star},j}^{(q-1,p-1)}(t)
}{\lambda_{n^{\star},\ell^{\star},j} - \lambda_{\tilde{n},\tilde{\ell}-1,j}} - \sqrt{\tilde{n}} \dfrac{
C_{\tilde{n}-1,\tilde{\ell}+1,j} ^{(q-1,p)}(t) - D_{n^{\star},\ell^{\star},j}^{(q-1,p)} (t)
}{\lambda_{n^{\star},\ell^{\star},j} - \lambda_{\tilde{n}-1,\tilde{\ell}+1,j}}
& \quad  \mbox{if } \tilde{\ell}\!>\!0
\\
\, \\
- \sqrt{\tilde{n}} \dfrac{
C_{\tilde{n}-1,\tilde{\ell}-1,j} ^{(q-1,p-1)}(t) - D_{n^{\star},\ell^{\star},j}^{(q-1,p-1)}(t)
}{\lambda_{n^{\star},\ell^{\star},j} - \lambda_{\tilde{n}-1,\tilde{\ell}-1,j}} - \sqrt{\tilde{n}} \dfrac{
C_{\tilde{n}-1,\tilde{\ell}+1,j} ^{(q-1,p)}(t) - D_{n^{\star},\ell^{\star},j}^{(q-1,p)} (t)
}{\lambda_{n^{\star},\ell^{\star},j} - \lambda_{\tilde{n}-1,\tilde{\ell}+1,j}}
& \quad \mbox{if } \tilde{\ell}\!=\!0 \\
\, \\
- \sqrt{\tilde{n}} \dfrac{
C_{\tilde{n}-1,\tilde{\ell}-1,j} ^{(q-1,p-1)}(t) - D_{n^{\star},\ell^{\star},j}^{(q-1,p-1)}(t)
}{\lambda_{n^{\star},\ell^{\star},j} - \lambda_{\tilde{n}-1,\tilde{\ell}-1,j}} + 
\sqrt{\tilde{n} \!-\! \tilde{\ell}} \dfrac{
C_{\tilde{n},\tilde{\ell}+1,j} ^{(q-1,p)}(t) - D_{n^{\star},\ell^{\star},j}^{(q-1,p)} (t)
}{\lambda_{n^{\star},\ell^{\star},j} - \lambda_{\tilde{n},\tilde{\ell}+1,j}}
& \quad  \mbox{if } \tilde{\ell}\!<\!0
\end{array}
\right. 
\end{align}
where $\tilde{\ell} = \ell^{\star} - q + 2p$ and $\tilde{n} = n^{\star} + (q-|\tilde{\ell}|+|\ell^{\star}|)/2$
and the initial conditions are given by
\begin{align}
D_{n^{\star}\!,\ell^{\star},j}^{(1,1)}(t) = t \, C_{n^{\star}\!,\ell^{\star},j}^{(0,0)} (t)
\left\lbrace \!\!
\begin{array}{lr}
\sqrt{n^{\star} \!+\! \ell^{\star} \!+\! 1}
& \;  \mbox{if } \ell^{\star} \!\geq\!0
\\
\, \\
-\sqrt{n^{\star} \!+\! 1}
& \; \mbox{if } \ell^{\star}  \!<\!0 \\
\end{array}
\right. 
\end{align}
\begin{align}
D_{n^{\star}\!,\ell^{\star},j}^{(1,0)}(t) = t \, C_{n^{\star}\!,\ell^{\star},j}^{(0,0)} (t) 
\left\lbrace \!\!
\begin{array}{lr}
-\sqrt{n^{\star} \!+\! 1}
& \;  \mbox{if } \ell^{\star} \!>\!0
\\
\, \\
\sqrt{n^{\star}  \!-\! \ell^{\star} \!+\! 1}
& \; \mbox{if } \ell^{\star}  \!\leq\!0 \\
\end{array}
\right. 
\end{align}
and
$D_{n^{\star}\!,\ell^{\star},j}^{(q,p)}(t) = 0$ if $q<1$ or $p<0$ or $p>q$.
However, the previous scheme does not solve the numerical problem if along the chain of $(n',\ell')$ pairs connecting $(n^{\star},\ell^{\star})$ to $(n,\ell)$ there is one or more $(n',\ell')$ such that also $\lambda_{n',\ell',j} \rightarrow \lambda_{n^{\star}\!,\ell^{\star}\!,j}$.
Since it is clear from Eq.~\eqref{eq:recursion_Cs_sup} that the functions $C$'s never display a singular behavior, also this problem can be solved by finding new integrated recursive relations whose complexity increases with the number of states having the same eigenvalue along the chain of pairs connecting $(n^{\star},\ell^{\star})$ to $(n,\ell)$.
However, this goes beyond the scope of the present work.

\clearpage

\begin{center}
{\large \textbf{Moments and correlation functions}}
\end{center}

Before calculating the moments and the correlation functions, lets derive some usefull relations starting from
\begin{align}
\psi_{n,\ell,j}(\vec{r},\vartheta) =  \sqrt{\frac{n!}{(n+|\ell|)!}}  \;  \left(\frac{r}{d\sqrt{2}} \right)^{|\ell|} \, \text{L}_n^{|\ell|}\! \left(\frac{r^2}{2d^2} \right)
e^{i \ell \varphi } e^{i (j-\ell) \vartheta}  \; ,
\end{align}
it is easy to show that
\begin{align}
r^2 = 2d^2 \left[   \psi_{0,0,0}(\vec{r},\vartheta)^* -  \psi_{1,0,0}(\vec{r},\vartheta)^* \right] \; ,
\end{align}
\begin{align}
r^4 = 4 d^4  \left[  2 \psi_{0,0,0}(\vec{r},\vartheta)^* - 4 \psi_{1,0,0}(\vec{r},\vartheta)^* + 2 \psi_{2,0,0}(\vec{r},\vartheta)^* \right] \; ,
\end{align}

\begin{align}
\cos \vartheta = \dfrac{1}{2} \left[   \psi_{0,0,-1}(\vec{r},\vartheta)^* +  \psi_{0,0,1}(\vec{r},\vartheta)^* \right] \; ,
\end{align}
\begin{align}
r \cos \varphi =  \dfrac{d}{\sqrt{2}}  \left[   \psi_{0,-1,-1}(\vec{r},\vartheta)^* +  \psi_{0,1,1}(\vec{r},\vartheta)^* \right] \; ,
\end{align}
\begin{align}
r \cos \varphi \cos \vartheta = \dfrac{d}{2\sqrt{2}} \left[   \psi_{0,-1,-2}(\vec{r},\vartheta)^* +  \psi_{0,-1,0}(\vec{r},\vartheta)^* + \psi_{0,1,0}(\vec{r},\vartheta)^* +  \psi_{0,1,2}(\vec{r},\vartheta)^* \right] \; ,
\end{align}
\begin{align}
r^2 \cos (2\varphi) = \sqrt{2} d^2 \left[   \psi_{0,-2,-2}(\vec{r},\vartheta)^* + \psi_{0,2,2}(\vec{r},\vartheta)^* \right] \; ,
\end{align}
and so on. 
Now we can easily calculate some paradigmatic moments and correlation functions. 
In particular, the positional autocorrelation function reads
\begin{align}
\langle x(t) x(0) \rangle = & \int \diff \vec{r}_0 \int_0^{2\pi} \diff \vartheta_0 \, r_0 \cos \varphi_0 \, p^{\text{eq}}(\vec{r}_0,\vartheta_0)
\sum_{n,\ell,j} M_{n,\ell,j}^{\infty} \, \psi_{n,\ell,j}(\vec{r}_0,\vartheta_0) \nonumber \\
& \times \int \diff \vec{r} \int_0^{2\pi} \diff \vartheta \, r \cos \varphi \, p^{\text{eq}}(\vec{r},\vartheta)
\sum_{n',\ell',j'} M_{n',\ell',j'}(\vec{r}_0,\vartheta_0,t) \, \psi_{n',\ell',j'}(\vec{r},\vartheta) \; ,
\end{align}
where the factors $M_{n,\ell,j}^{\infty} = \lim_{t\rightarrow \infty} M_{n,\ell,j}(\vec{r},\vartheta,t)$ do not depend on time and position anymore.
Thus, using the orthogonality property
\begin{align}
\int \diff \vec{r} \int_0^{2\pi} \diff \vartheta \, p^{\text{eq}}(\vec{r},\vartheta) \, \psi_{n',\ell',j'}(\vec{r},\vartheta) \, \psi_{n,\ell,j}(\vec{r},\vartheta)^* = \delta_{n,n'} \delta_{\ell,\ell'} \delta_{j,j'} \; ,
\end{align}
we have
\begin{align}
\langle x(t) x(0) \rangle = & \dfrac{d}{\sqrt{2}} \int \diff \vec{r}_0 \int_0^{2\pi} \diff \vartheta_0 \, r_0 \cos \varphi_0 \, p^{\text{eq}}(\vec{r}_0,\vartheta_0)
\sum_{n,\ell,j} M_{n,\ell,j}^{\infty} \, \psi_{n,\ell,j}(\vec{r}_0,\vartheta_0) \nonumber \\
& \times  \left[   M_{0,-1,-1}(\vec{r}_0,\vartheta_0,t) +  M_{0,1,1}(\vec{r}_0,\vartheta_0,t) \right] \nonumber \\
= &  \int \diff \vec{r}_0 \int_0^{2\pi} \diff \vartheta_0 \, r_0 \cos \varphi_0 \, p^{\text{eq}}(\vec{r}_0,\vartheta_0)
\sum_{n,\ell,j} M_{n,\ell,j}^{\infty} \, \psi_{n,\ell,j}(\vec{r}_0,\vartheta_0) \nonumber \\
& \times  \left[  r_0 \cos \varphi_0 \, e^{-t/\tau} - \text{Pe} \, d  \cos \vartheta_0 \dfrac{e^{-t/\tau} - e^{-D_{\text{rot}}t}}{1 -  D_{\text{rot}}\tau} \right] \nonumber \\
= & \, d^2 \left[ M_{0,0,0}^{\infty} - M_{1,0,0}^{\infty} \right] e^{-t/\tau} + \dfrac{d^2}{\sqrt{2}} \left[ M_{0,-2,-2}^{\infty} + M_{0,2,2}^{\infty} \right] e^{-t/\tau} \nonumber \\
& - \text{Pe} \dfrac{d^2}{2 \sqrt{2}}  \left[ M_{0,-1,-2}^{\infty} + M_{0,-1,0}^{\infty} + M_{0,1,0}^{\infty} + M_{0,1,2}^{\infty} \right] \dfrac{e^{-t/\tau} - e^{-D_{\text{rot}}t}}{1 - D_{\text{rot}}\tau} \; ,
\end{align}
where we used
\begin{align}
r^2 \cos^2 \varphi = \dfrac{r^2}{2} [ 1 +  \cos(2\varphi) ] \; .
\end{align}
Since $M_{0,0,j}^{\infty} = \delta_{j,0}$ and since the iterative scheme provided in the previous section does not change $j$, it follows that if $j\neq 0$ then $M_{n,\ell,j}^{\infty} = 0$ for each $n$ and $\ell$.
Thus
\begin{align}
\langle x(t) x(0) \rangle = & \, d^2 \left[ M_{0,0,0}^{\infty} - M_{1,0,0}^{\infty} \right] e^{-\mu k t} - \text{Pe} \dfrac{d^2}{2\sqrt{2}} \left[  M_{0,-1,0}^{\infty} + M_{0,1,0}^{\infty} \right] \dfrac{e^{- t/\tau} - e^{-D_{\text{rot}}t}}{1 - D_{\text{rot}}\tau} \nonumber \\
= & d^2 \left[ 1 +  \dfrac{\text{Pe}^2}{2(1 + \tau D_{\text{rot}})} \right] e^{-t/\tau} - \text{Pe}^2 \dfrac{d^2}{2 (1+ D_{\text{rot}}\tau )} \dfrac{e^{-t/\tau} - e^{-D_{\text{rot}}t}}{1 - D_{\text{rot}}\tau}\; .
\end{align}
Since the harmonic well is isotropic and given the integration over the initial conditions we also have
\begin{align}
\langle y(t) y(0) \rangle = \langle x(t) x(0) \rangle \; .
\end{align}

Using the above results, one also evaluates the mean square displacement
\begin{align}
\langle [\vec{r}(t)-\vec{r}(0)]^2 \rangle = \langle r^2(t) \rangle - 2 \langle \vec{r}(t) \cdot \vec{r}(0) \rangle + \langle r^2(0) \rangle \; .
\end{align}
Noting that because of the average over the initial condition  $\langle r^2(t) \rangle = \langle r^2(0) \rangle$ we just need to calculate
\begin{align}
\langle r^2(0) \rangle = & \int \diff \vec{r}_0 \int_0^{2\pi} \diff \vartheta_0 \, r_0^2 \, p^{\text{eq}}(\vec{r}_0,\vartheta_0)
\sum_{n,\ell,j} M_{n,\ell,j}^{\infty} \, \psi_{n,\ell,j}(\vec{r}_0,\vartheta_0)  \nonumber \\
= & \, 2d^2 \left[  M_{0,0,0}^{\infty} - M_{1,0,0}^{\infty} \right] =  2d^2 + \dfrac{\text{Pe}^2 \, d^2}{1 + D_{\text{rot}}\tau} \; ,
\end{align}
to finally obtain
\begin{align}
\langle [\vec{r}(t)-\vec{r}(0)]^2 \rangle = 4d^2 \left( 1-e^{-\mu k t}\right) + 2 \, \text{Pe}^2 d^2  \left[  \dfrac{1-e^{- t/\tau}}{(1 + D_{\text{rot}}\tau) } + \dfrac{e^{-t/\tau}-e^{- D_{\text{rot}} t}}{1 - (D_{\text{rot}}\tau)^2 } \right] \; .
\end{align}

\medskip

It is also interesting being able to evaluate the moments given a specific initial condition. For instance
\begin{align}
\langle r^2(t) \rangle_{r_0,\vartheta_0} = & \int_0^{\infty} \diff \vec{r} \int_0^{2\pi} \diff \vartheta \; r^2 \, p^{\text{eq}}(\vec{r},\vartheta)
\sum_{n,\ell,j} M_{n,\ell,j}(\vec{r}_0,\vartheta_0,t) \, \psi_{n,\ell,j}(\vec{r},\vartheta)
\nonumber \\
= & 2 d^2 \left[  M_{0,0,0}(\vec{r}_0,\vartheta_0,t) - M_{1,0,0}(\vec{r}_0,\vartheta_0,t) \right]  \nonumber \\
= & 2d^2 \left[ 1- e^{-2 t/\tau} \right] + r_0^2 e^{-2 t/\tau} 
+2 \, \text{Pe} \, d \,  r_0 \cos( \varphi_0 - \vartheta_0)  \, \dfrac{e^{-t/\tau -D_{\text{rot}}t} -e^{-2 t/\tau}}{1 - D_{\text{rot}}\tau} \nonumber \\
& + \text{Pe}^2 \, d^2 \,  \left[  \dfrac{1}{ 1 + D_{\text{rot}}\tau}  - \dfrac{2e^{-t/\tau - D_{\text{rot}}t}}{ 1 - (D_{\text{rot}}\tau)^2} + \dfrac{e^{-2 t/\tau}}{ 1 -D_{\text{rot}}\tau} \right] \; ,
\end{align}
and
\begin{align}
\langle x \rangle_{r_0,\vartheta_0}  = \frac{d}{\sqrt{2}}  \left[ M_{0,1,1}(\vec{r}_0,\vartheta_0,t) \!+\! M_{0,1,-1}(\vec{r}_0,\vartheta_0,t) \right] 
 = x_0 e^{-t/\tau} + \text{Pe} \,  d \, \cos(\vartheta_0) \dfrac{e^{-D_{\text{rot}}t} - e^{-t/\tau}}{1 - D_{\text{rot}}\tau} \; ,
\end{align}
where $\langle \bullet \rangle_{r_0,\vartheta_0}$ denotes an average at fixed initial conditions.

\end{document}